\documentclass[preprint,12pt]{elsarticle}
\usepackage{amssymb}
\usepackage{graphicx}
\usepackage{subfigure}

\journal{Safety Science}

\begin{document}
\begin{frontmatter}

%% Title, authors and addresses

%% use the tnoteref command within \title for footnotes;
%% use the tnotetext command for the associated footnote;
%% use the fnref command within \author or \address for footnotes;
%% use the fntext command for the associated footnote;
%% use the corref command within \author for corresponding author footnotes;
%% use the cortext command for the associated footnote;
%% use the ead command for the email address,
%% and the form \ead[url] for the home page:
%%
%% \title{Title\tnoteref{label1}}
%% \tnotetext[label1]{}
%% \author{Name\corref{cor1}\fnref{label2}}
%% \ead{email address}
%% \ead[url]{home page}
%% \fntext[label2]{}
%% \cortext[cor1]{}
%% \address{Address\fnref{label3}}
%% \fntext[label3]{}

\title{An improved vulnerability index of complex networks based on fractal dimension}

%% use optional labels to link authors explicitly to addresses:
%% \author[label1,label2]{<author name>}
%% \address[label1]{<address>}
%% \address[label2]{<address>}

\author{\footnotesize Li Gou$^a$, Bo Wei$^a$, Rehan Sadiq$^b$, Sankaran Mahadevan$^c$, Yong Deng$^{a,c,}$\footnote{Corresponding author: Yong Deng, School of
Computer and Information Science, Southwest University,
Chongqing, 400715, China. Email address: ydeng@swu.edu.cn; prof.deng@hotmail.com}}

\address{$^a$School of Computer and Information Science,\\ Southwest University, Chongqing 400715, China\\
$^b$School of Engineering, University of British Columbia Okanagan,\\
 3333 University Way, Kelowna, BC, Canada V1V 1V7\\
$^c$School of Engineering, Vanderbilt University, Nashville, TN 37235, USA}

\begin{abstract}
With an increasing emphasis on network security, much more attention has been attracted to the vulnerability of complex networks. The multi-scale evaluation of vulnerability is widely used since it makes use of combined powers of the links' betweenness and has an effective evaluation to vulnerability. However, how to determine the coefficient in existing multi-scale evaluation model to measure the vulnerability of different networks is still an open issue. In this paper, an improved model based on the fractal dimension of complex networks is proposed to obtain a more reasonable evaluation of vulnerability with more physical significance. Not only the structure and basic physical properties of networks is characterized, but also the covering ability of networks, which is related to the vulnerability of the network, is taken into consideration in our proposed method. The numerical examples and real applications are used to illustrate the efficiency of our proposed method.

\end{abstract}

\begin{keyword}
vulnerability\sep fractal dimension\sep complex networks
%% keywords here, in the form: keyword \sep keyword

%% MSC codes here, in the form: \MSC code \sep code
%% or \MSC[2008] code \sep code (2000 is the default)

\end{keyword}

\end{frontmatter}

%%
%% Start line numbering here if you want
%%
% \linenumbers

%% main text
\section{Introduction}\label{section one}
Complex network are widely used to model the structure of many complex systems in nature and society \cite{kim2012analysis,wang2009complex,tang2011detecting,qi2010efficiency,zhang2013self,holme2002attack,MonfaredPA2014}. An open issue is how to assess the vulnerability of complex networks \cite{holmgren2006using,boccaletti2007multiscale,zhang2012attack,wang2013vulnerability}, whose main objective is to understand, predict, and even control the behavior of a networked system under vicious attacks or any types of dysfunctions \cite{boccaletti2007multiscale,zhang2013route}.

Different approaches to characterize network vulnerability and robustness have recently been proposed, which can be grouped into two types broadly \cite{mishkovski2011vulnerability,ouyang2014correlation}. The first type of approach is related to structural robustness \cite{mishkovski2011vulnerability,albert2004structural,albert2002statistical}: how topological properties of networks are affected by the removal of a finite number of vertexes or/and links, such as the degree distribution, the network connectivity level, the size of largest component, the average geodesic length and etc. The second type of method concerns dynamical robustness \cite{holme2002attack,motter2002cascade,crucitti2004model,wang2009cascade,wang2013robustness}. The removal of a vertex or link will cause the flow to redistribution with the risk that some other vertexes or links may be overloaded, which can cause a sequence of failures and even threaten the global stability. Such behavior is called cascading failures \cite{mishkovski2011vulnerability,wang2009vulnerability,wang2013improving,wang2011robustness}.

One of the mostly used methods is proposed by Boccaletti \emph{et.al } \cite{boccaletti2007multiscale}. They construct a multi-scale evaluation model of vulnerability, which makes use of combined powers of the links' betweenness. Due to the simplicity and efficiency, this method is heavily studied \cite{mishkovski2011vulnerability}. One limitation of original model is that it cannot discriminate two different networks in some situations. To solve this problem, a coefficient $p$ is introduced to improve the original model. However, a straight problem is that how to determine the coefficient $p$. The method to determine the coefficient $p$ in Boccaletti \emph{et.al }' work is very complicated and lack of physical significance.

The main motivation of our work is that we believe that this coefficient should be determined by the network itself. To address this issue, we take the fractal dimension of complex network into consideration. The dimension of complex networks is one of the most fundamental quantities to characterize its structure and basic physical properties \cite{daqing2011dimension,shanker2007graph,wei2013new}. One has proved that the network dimension is a key concept to understand not only network topology. But also dynamical process on networks, such as diffusion and critical phenomenon including percolation, which is also used to characterize the vulnerability of network. Box covering algorithm \cite{song2005self,song2007calculate,wei2013box} are one of the typical ways to calculate the fractal dimension \cite{shanker2008algorithms}. In short, fractal dimension is a key parameter to represent the characters of the network. Based on this idea, we propose that the dimension of the network has a significant relation with network vulnerability in this paper.

This paper is organized as follows. Section 2 introduces the preliminaries. In Section 3 we calculate the vulnerability of some networks using the proposed method. In Section 4 we compare the proposed method with the existing methods in other papers by calculating network vulnerability. Finally, we summarize our results in Section 5.

%%discuss and testify the relationship between network vulnerability and fractal dimension, and use the proposed method to calculate the vulnerability of some representative networks.%%

\section{Preliminaries}\label{section two}
In this section, we introduce Boccaletti \emph{et.al }'s model\cite{boccaletti2007multiscale} and three other methods \cite{holme2002attack,holmgren2006using,mishkovski2011vulnerability}.
In general, the complex networks can be represented by an undirected and unweighted graph $G={(V, E)}$, where $V$ is the set of vertices and $E$ is the set of edges. Each edges connects exactly one pari of vertices, and a vertex-pair can be connected by maximally one edge, i.e. loop is not allowed.

In Boccaletti \emph{et.al }'s work \cite{boccaletti2007multiscale}, the original method to evaluate the vulnerability is represented by the average edge betweenness, which is defined as:
\begin{equation}\label{average edge betweenness}
  b_1{(G)}=\frac{1}{|E|}\sum_{l\in E}{b_l},
\end{equation}
where $|E|$ is the number of the edges, and $b_l$ is the edge betweenness of the edge $l$, define as:
\begin{equation}\label{edge betweenness}
  b_l=\sum_{{j,k} \in V} {\frac{n_{jk}{(l)}}{n_{jk}}},
\end{equation}
where $n_{jk}{(l)}$ is the number of geodesics(shortest path) from $j$ to $k$ that contain the link $l$, and $n_{jk}$ is the total number of geodesics from $j$ to $k$.

\begin{figure*}[htbp]
\begin{center}
\includegraphics[width=10cm]{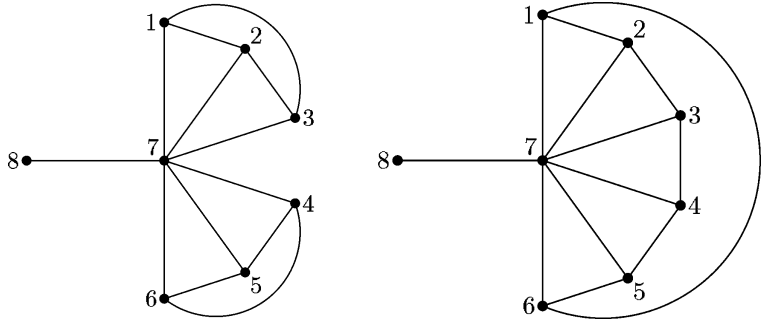}
\caption{The ``bat" graph $G$ and the ``umbrella" graph $G'$\cite{boccaletti2007multiscale}.}\label{comparison_figure}
%(d) the US power-grid. (f)network of e-mail interchanges.
\vspace{0.2in}
\end{center}
\end{figure*}

However, this evaluation of $b_1{(G)}$ gives no relevant new information about the vulnerability of the network. For example, two networks referred in \cite{boccaletti2007multiscale} shown in Fig. \ref{comparison_figure} can't be distinguished using this method. By evaluating the vulnerability according to Eq. \ref{average edge betweenness}, one gets $b_1{(G)}=b_1{(G')}=43/13$. It's absolute that the ``bat" graph $G$ is more vulnerable than the ``umbrella" graph $G'$, but Eq. \ref{average edge betweenness} gives the same evaluation result.

In order to overcome the original method's limitation of failing to distinguish some networks, the coefficient $P$ was introduced to evaluate vulnerability of complex network, which is called multi-scale evaluation of vulnerability \cite{boccaletti2007multiscale} and shown as below:
\begin{equation}\label{vulnerability}
  b_p{(G)}=(\frac{1}{|E|}\sum_{l\in E}{b_l^p})^{\frac{1}{|p|}}
\end{equation}
for each value of $p>0$. If we want to compare two networks $G$ and $G^{'}$, first computes $b_1$. If $b_1{(G)}<b_1{(G^{'})}$, then $G$ is more robust than $G^{'}$. On the other hand, if $b_1{(G)}=b_1{(G^{'})}$ then one takes $p>1$ and computes $b_p$ until $b_p{(G)}\neq {b_p{(G^{'})}}$.

To get the coefficient $p$, Boccaletti \emph{et.al } define a relative function of $p$ like:
 \begin{equation}\label{relative function}
  f(p)=(b_p{(G)}-b_p{(G^{'})})/b_p{(G)}
\end{equation}
 The coefficient $p$ is obtained when the function has a maximal value. For the more detailed information to determine the coefficient $p$, refer \cite{boccaletti2007multiscale}. It's clear that, the coefficient $p$'s definition is complicated and lack of physical significance. The coefficient $p$ should reflect the complex network itself.

For the sake of comparison, three other methods to calculate vulnerability are described as follows. The first method is the average inverse geodesic length $l^{-1}$ \cite{holme2002attack}:
\begin{equation}\label{inverse}
  l^{-1}=\langle{\frac{1}{d{(v,w)}}}\rangle\equiv{\frac{1}{N(N-1)}\sum{\sum{\frac{1}{d{(v,w)}}}}}.
\end{equation}
where $d(v,w)$ is the length of the geodesic between $v$ and $w$ ($v,w\in{V}$). The larger $l^{-1}$ is, more robust the network is.

The second method is the largest component size $LCS$ ($0<LCS<1$) \cite{holmgren2006using}, which quantifies the number of nodes in the largest connected subgraph and defined as follows:
\begin{equation}\label{LCS}
  LCS=\frac{N_s}{N}
\end{equation}
where $N_s$ is the size of the largest connected subgraph.

And the third method is the normalized average edge betweenness $b_{nor}{(G)}$ \cite{mishkovski2011vulnerability}, which is on the base of the Eq. \ref{vulnerability} while $p=1$ and is defined as:
\begin{equation}\label{normalized betweenness}
  b_{nor}(G)=\frac{b_1{(G)}-b_1{(G_{complete})}}{b_1{(G_{path})}-b_1{(G_{complete})}}=\frac{b_1{(G)}-1}{\frac{N(N+1)}{6}-1}.
\end{equation}
where $G_{complete}$ is a complete graph and $G_{path}$ is a path graph.

\section{Proposed vulnerability model}\label{section three}

\begin{figure*}[htbp]
\begin{center}
\includegraphics[width=8cm]{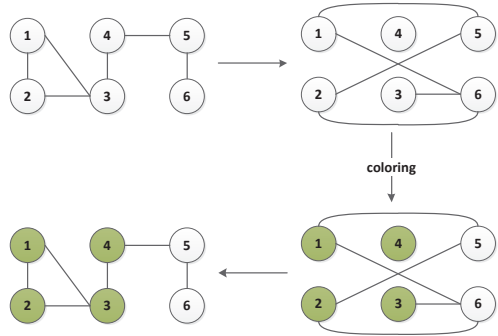}
\caption{Illustration of the box-covering algorithms. Starting from $G$ (upper left panel), a dual network $G'$ (upper right panel) was constructed for a given box size (here $l_B= 3$), where two nodes are connected if they are at a distance $l\geq{l_B}$. A greedy algorithm was used for vertex colouring in $G'$, which is then used to determine the box covering in $G$, as shown in the plot\cite{song2007calculate}.}\label{boxcovering_figure}
%(d) the US power-grid. (f)network of e-mail interchanges.
\vspace{0.2in}
\end{center}
\end{figure*}

In this section, the proposed method is detailed. As mentioned in introduction section, We think that the coefficient $p$ should be determined by the network itself. In addition, this coefficient $p$ should also has the direct relation to the vulnerability of this network. In our opinion, the fractal dimension of the network is a promising alternative. For a given network $G$ and box size $l_B$, a box is a set of nodes where all distances $l_{ij}$ between any two nodes $i$ and $j$ in the box are smaller than $l_B$. The minimum number of boxes required to cover the entire network is denoted by $N_B$. The detailed illustration referred in \cite{song2007calculate} of the calculation of the fractal dimension is given in Fig. \ref{boxcovering_figure}.  The fractal dimension or box dimension $d_B$ calculated with the box covering algorithm is given as follows \cite{song2005self,song2007calculate}:
\begin{equation}\label{fractal dimension}
  N_B\approx{l_B^{-d_B}}
\end{equation}

It is very known that the fractal dimension can characterize the network structure and basic physical properties which reflects the covering ability. For a given network, the higher the fractal dimension, the higher the covering ability, which means that there are more edges between the nodes in this network. We also know that given certain nodes in the network, the more edges, the more robust of this network. As a result, the fractal dimension not only reflects the characters of the network structure, but also partially reflects the vulnerability of the network. According to this idea, we use the fractal dimension to redefine $p$. So the proposed method to calculate network vulnerability is given as follows:
\begin{equation}\label{vulnerability with dimension}
  V_{d_B}{(G)}=(\frac{1}{|E|}\sum_{l\in E}{b_l^{d_B}})^{\frac{1}{|d_B|}}
\end{equation}
where $d_B$ is the fractal dimension of the complex networks.

We apply our method to six networks to calculate the vulnerability index. Two are synthetic networks, Erd\H{o}s-R\'{e}nyi(ER) random networks \cite{erdos1960evolution} and Barab\'{a}si-Albert(BA) model of scale-free networks \cite{barabasi1999emergence}. Four are real networks: US airport networks \cite{colizza2007reaction},  network of e-mail interchanges \cite{Guimera2003email}, protein-protein interaction network \cite{ppi_data} and German highway system \cite{kaiser2004spatial}.
%Owing to the big difference in the vertices of these networks, we classify these networks into 2 groups: ER, BA, EI, AP, PPI, GH and IR, GH.

\begin{figure*}[htbp]
\begin{center}
\subfigure[]{
\includegraphics[width=6cm]{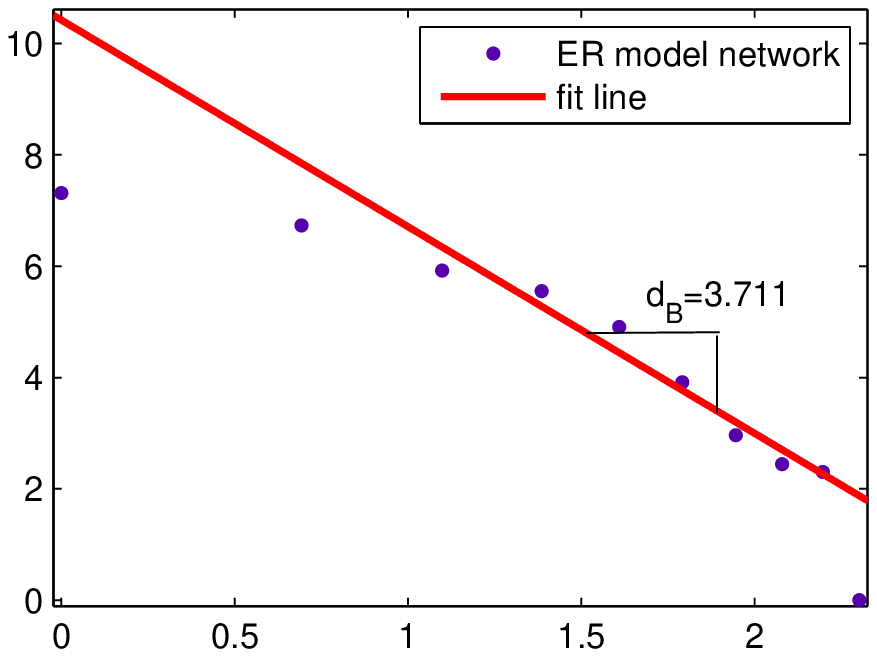}
}
\subfigure[]{
\includegraphics[width=6cm]{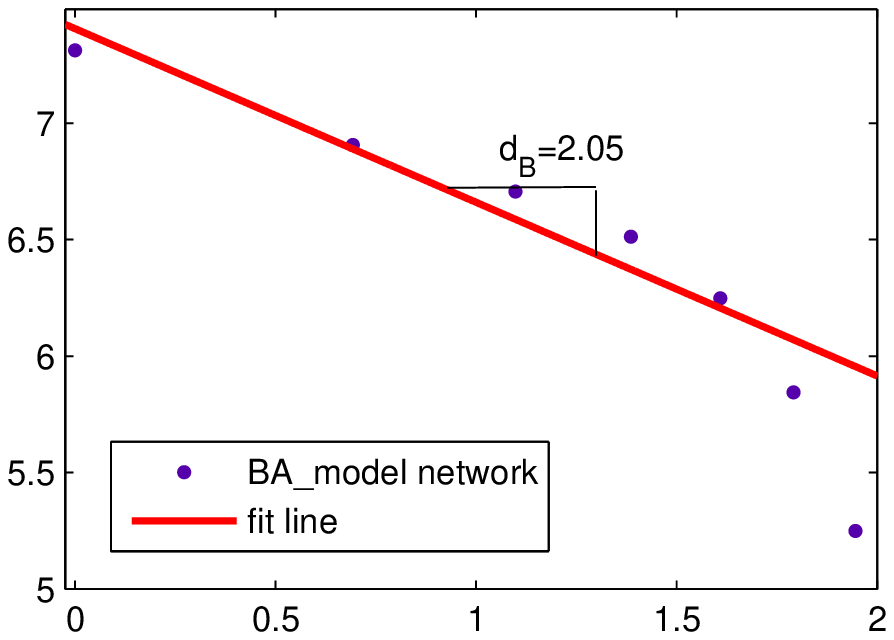}
}
\subfigure[]{
\includegraphics[width=6cm]{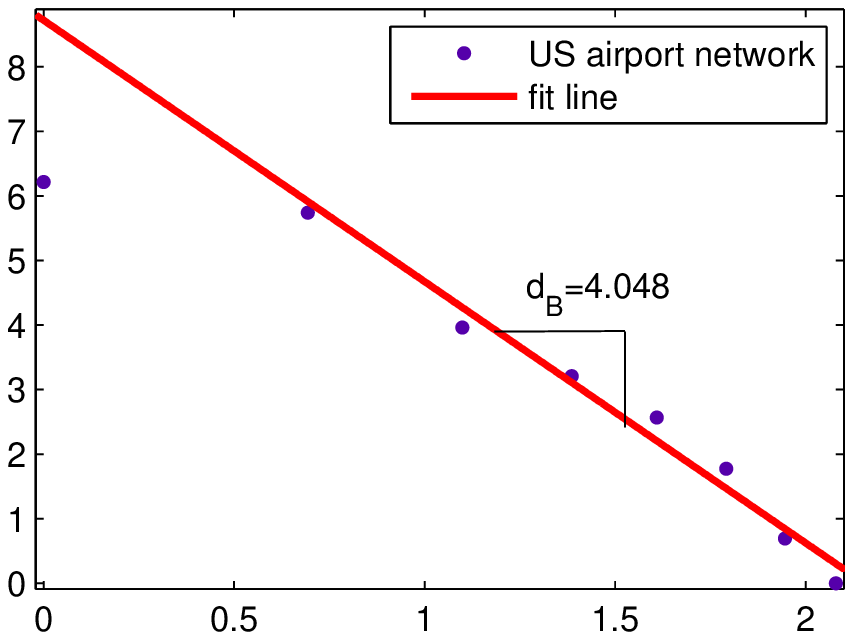}
}
\subfigure[]{
\includegraphics[width=6cm]{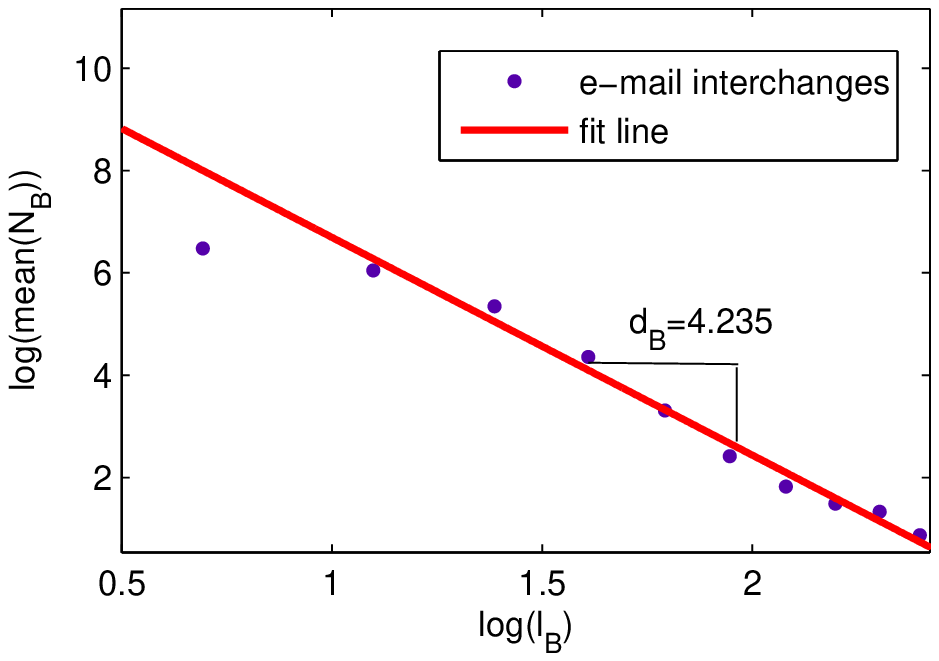}
}
\subfigure[]{
\includegraphics[width=6cm]{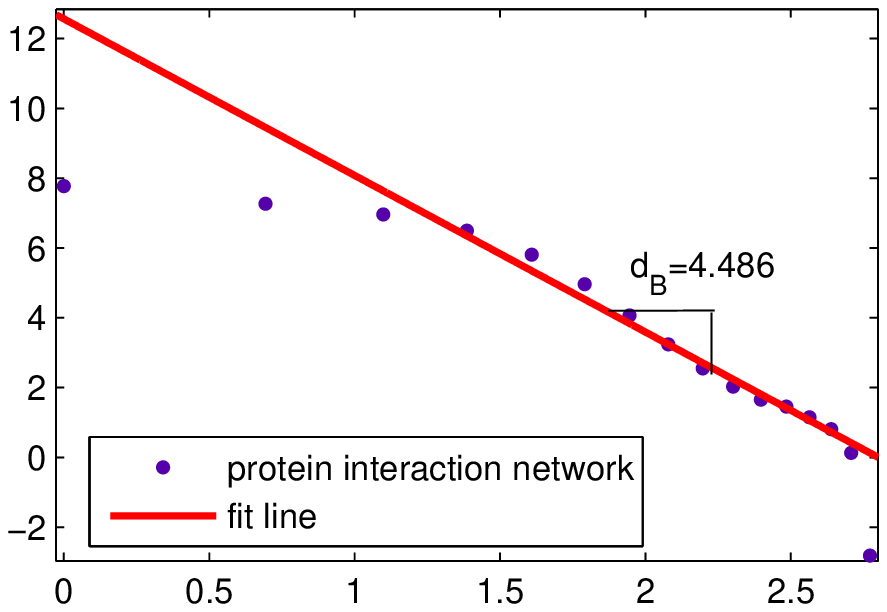}
}
\subfigure[]{
\includegraphics[width=6cm]{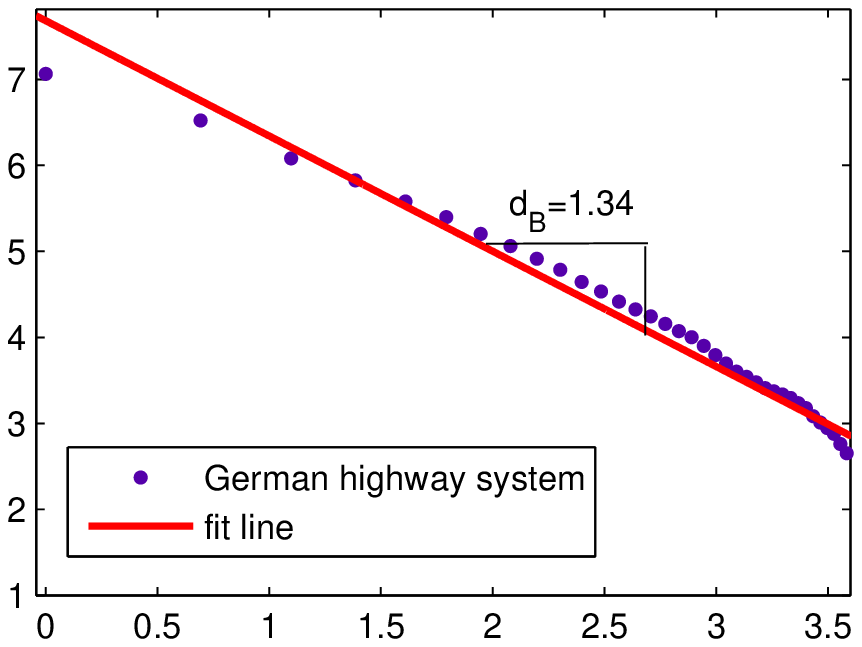}
}
%\subfigure[]{
%\includegraphics[width=6cm]{router.eps}
%}
%\subfigure[]{
%\includegraphics[width=6cm]{power_grid.eps}
%}
\caption{The $N_B$ versus $l_B$ of some complex networks obtained in a log-log scale: (a) the ER network with the size $N=1500$, the average degree $<k>=6$. (b) the BA network with $N=1500$, the average degree $<k>=4.8$. (c) US airport network. (d) network of e-mail interchanges. (e) protein-protein interaction network. (f) German highway system. The vertical ordinate of every subplot is the mean value of $N_B$ for 100 times, and the horizonal ordinate represents the box size $l_B$. The absolute value of the slope is the fractal dimension.
}\label{dimension_figure}
%(d) the US power-grid. (f)network of e-mail interchanges.
\vspace{0.2in}
\end{center}
\end{figure*}

%network of Internet route level (IR)\cite{spring2004measuring}, the high-voltage power grid of the United States of America (PG)\cite{watts1998collective},
The vulnerability of these networks are calculated according to the follow steps:

 (1) calculate the fractal dimension $d_B$ of these networks above using box-covering algorithm \cite{song2005self,song2007calculate}, i.e. Eq. \ref{fractal dimension}. The results are illustrated in Fig. \ref{dimension_figure}.

(2) Calculate the average edge betweenness according to Eq. \ref{edge betweenness}, and normalized by $\frac{N(N-1)}{2}$.

(3) Calculate the vulnerability $V_{d_B}$ in accordance with Eq. \ref{vulnerability with dimension}.

 Table \ref{table1} shows the result. The larger the $V_{d_B}$, the more vulnerable the network. So the results illustrate the order of vulnerability $GH >AP > PPI > BA > EI > ER $, and the robustness of networks correspond to the inverse order.

%Owing to the big difference in the vertices of these networks, we classify these networks into 2 groups:
\begin{table}[htbp]
{\footnotesize
\caption{General characteristics of several complex networks. For each network we list the number of nodes $N$, the average degree $<k>$, the fractal dimension $d_B$, and the vulnerability $V_{d_B}$ obtained by the proposed method. ER, BA, AP, EI, PPI and GH denote the ER network, the BA network, US airport network, network of e-mail interchanges, protein-protein interaction network and the German highway system.  }\label{table1}
\begin{tabular*}{\columnwidth}{@{\extracolsep{\fill}}@{~~}cccccc@{~~}}
\hline
network  & $N$ & $<k>$ & $d_B$ & $V_{d_B}$  \\
\hline
   ER  & 1500 & 6 &3.711  &0.0011\\
   BA  & 1500 &  4.8   &2.05  &0.0014\\
   AP  & 500 & 11.9   &4.048  &0.0079\\
   EI  & 1134 &  9.6   &4.235   &0.0013\\
   PPI  & 2375 & 9.8 &4.486 &0.0037\\
   GH  & 1168 & 2.1 & 1.34 & 0.0184\\
  % IR  & 5022 & 2.5 &3.22 &0.0111 \\
%   PG  & 4941 & 2.7 &6.728 &0.0859  \\
\hline
\end{tabular*}
}
\end{table}

\section{Comparison and Discussion}\label{section four}
In this section, to testify the correctness of the results obtained by the proposed method, three other methods presented in section \ref{section two} are applied to these networks to calculate the vulnerability, that is, the average inverse geodesic length $l^{-1}$, the largest component size $LCS$ and  the normalized average edge betweenness $b_{nor}(G)$. All three methods can reflect static topological properties of networks, in order to get the vulnerability reflecting the dynamical overall characteristics of networks, we apply the RB attack strategy \cite{holme2002attack} to networks when calculating them. RB attack strategy means that one should remove the node which has the highest betweenness value and recalculate the betweenness at every vertices-removing step. In this paper, $l^{-1}$, $LCS$, $b_{nor}(G)$ are computed after 1\% of vertices are removed. Table \ref{table2} shows the results.

\begin{table}[htbp]
{\footnotesize
\caption{The normalized average inverse geodesic length $\widetilde{l^{-1}}$, normalized largest component size $\widetilde{LCS}$ and the  average edge betweenness $b_{nor}{(G)}$ is computed after 1\% of vertices are removed.  All of them are normalized by the values of the initial networks.}\label{table2}
\begin{tabular*}{\columnwidth}{@{\extracolsep{\fill}}@{~~}cccccc@{~~}}
\hline
network  & $V_{d_B}$ & $\widetilde{l^{-1}}$ & $\widetilde{LCS}$ & $b_{nor}{(G)}$  \\
\hline
   ER  & 0.0011 &0.9788 &0.9886 &0.0666\\
   BA  & 0.0014 &0.8152 &0.9613 &0.4874\\
   AP  & 0.0079 &0.6259 &0.746  &-0.3563\\
   EI  & 0.0013 &0.9466 &0.9841 &0.1490\\
   PPI & 0.0037 &0.7681 &0.9175 &0.0912\\
   GH  & 0.0184 &0.5119 &0.9144 &0.7644\\
   %IR  & 0.0111 &0.0264 &0.1384 &-0.9863 \\
%   PG  & 0.0859 &0.2293 &0.2092 &-0.8923\\
\hline
\end{tabular*}
}
\end{table}
%$GH >AP > PPI > EI > ER > BA$

All the methods can give a rank about the vulnerability of these networks. The $\widetilde{l^{-1}}$ gives a order $ER > EI > BA > PPI > AP > GH$ about the robustness, and a robustness order $ER > EI > BA > PPI > GH > AP$ judging from $\widetilde{LCS}$, whereas $b_{nor}{(G)}$ ranks $GH > BA > EI > PPI > ER > AP$ in point of vulnerability. One can see that, The German highway system has the largest vulnerability and  for all the methods. The proposed method and the $\widetilde{l^{-1}}$ shows a completely identical order. So the proposed method is an effect way to quantify the network vulnerability.

\begin{figure*}[htbp]
\begin{center}
\includegraphics[width=8cm]{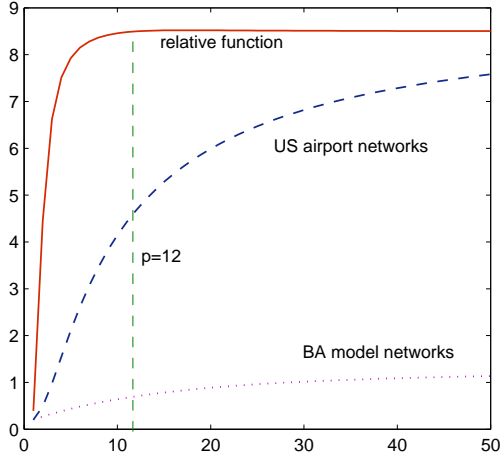}
\caption{$b_p$ for Barab\'{a}si-Albert(BA) model of scale-free networks (dot line) and US airport networks (dash line) as functions of $1\leq{p}\leq\infty$. $10(b_p{(AP)}-b_p{(BA)})/b_p{(AP)}$(solid line) as a relative function of $1\leq{p}\leq\infty$ has a unique maximum at $p=12$}\label{multi comparison}
%(d) the US power-grid. (f)network of e-mail interchanges.
\vspace{0.2in}
\end{center}
\end{figure*}

In a addition, the multi-scale model to calculate vulnerability proposed by Boccaletti \emph{et.al } are applied to these networks. A comparison of the proposed method and Boccaletti \emph{et.al }'s are illustrated in Table \ref{multi table} and Fig. \ref{multi comparison}.  As mentioned in section \ref{section two}, we should firstly compute $b_1{(G)}$ to judge if $p=1$ can distinguish these networks. Through computing, we found that $b_1{(BA)}=b_1{(AP)}=0.001$, which mean that the coefficient $p$ should be recalculated according to the relative function (Eq. \ref{relative function}). When the relative function has a maximal value, $p$ is obtained. We get $p=12$ for BA and AP networks, $b_p{(BA)}=0.0035,\quad b_p{(AP)}=0.0234$. Boccaletti \emph{et.al }'s method gives a order $PPI > EI > ER > BA > GH > AP$ about the robustness.

It's absolute that, the coefficient $p$ obtained by Boccaletti \emph{et.al }'s method is lack of physical meaning. Comparing rank orders obtained by these method, it's easy to found that the proposed method gives a more reasonable order and a more effective evaluation.
\begin{table}[htbp]
\centering
\small
\caption{\label{multi table}Some hypothetical scenarios to demonstrate the comparison of two methods.}
\begin{tabular}{lccccc}
\hline
network  &$p$($d_B$) of the & $V_{d_B}$ & $b_1{(G)}$  & $p$ of Boccaletti \emph{et.al }'s  & $b_p{(G)}$  \\&proposed method&& &method&\\
\hline
   ER  &3.711  &0.0011 &$9.3909e^{-004}$ &1  &$9.3909e^{-004}$\\
   BA  &2.05  &0.0014 &0.001            &12 &0.0035\\
   AP  &4.048  &0.0079 &0.001            &12 &0.0234\\
   EI  &4.235  &0.0013 &$6.6037e^{-004}$ &1  &$6.6037e^{-004}$\\
   PPI &4.486  &0.0037 &$4.3581e^{-004}$ &1  &$4.3581e^{-004}$\\
   GH  &1.34  &0.0184 &0.0156           &1  &0.0156\\
\hline
\end{tabular}
\end{table}

\section{Conclusions}\label{section five}
The coefficient $p$ used in the multi-scale model plays an important role in the vulnerability evaluation. How to determine the coefficient is still an open issue. The existing method is complex and lack of physical significance. To address this issue, an improved vulnerability index is proposed based on the fractal dimension of complex networks. The fractal dimension is one of the fundamental properties of complex networks, which can not only characterize the physical properties of networks, but also reflect the covering ability of networks. As a result, the new model has more meaning in physical aspect compared with existing methods. The numerical examples and real applications are used to illustrate the efficiency of our proposed method.

\section*{Acknowledgments}
The work is partially supported by National Natural Science Foundation of China (Grant No. 61174022), R$\&$ D Program of China (2012BAH07B01), National High Technology Research and Development Program of China (863 Program) (Grant No. 2013AA013801).

\bibliographystyle{elsarticle-num}
\bibliography{bibfile}
\end{document}